\documentclass[letterpaper]{JHEP}

\usepackage{epsfig}
\usepackage{graphicx}
\preprint{SUGP-00/6-1\\
 hep-th/0006117}

\title{Chern-Simons terms and the Three Notions of Charge}
\author{Donald Marolf\\Physics Department, Syracuse University, Syracuse,
         New York 13244}
\date{May, 2000}
\abstract{
In theories with Chern-Simons terms or modified Bianchi identities, 
it is useful to define three notions of either electric or magnetic
charge associated with a given gauge field.  A language for discussing these
charges is introduced and the properties of each charge are described.
`Brane source charge' is gauge invariant and localized but not conserved or quantized,
`Maxwell charge' is gauge invariant and conserved but not localized or quantized, while
`Page charge' conserved, localized, and quantized but not gauge invariant.
This provides a further perspective on the issue of charge quantization recently
raised by Bachas, Douglas, and Schweigert.  For the Proceedings of the
E.S. Fradkin Memorial Conference.
}
\begin{document}
%%%%%%%%%%%%%%%%%%%%%%%%%%%%%%%%%%%%%%%%%%%%%%%%%%%%%%
\section{Introduction}

One of the intriguing properties of supergravity theories is the presence
of Abelian Chern-Simons terms and their duals, the modified Bianchi identities, in
the dynamics of the gauge fields.  Such cases have the unusual feature that the equations
of motion for the gauge field are non-linear in the gauge fields
even though the associated gauge groups are Abelian.  For example, massless type IIA
supergravity contains a relation of the form
\begin{equation}
\label{MBid}
d\tilde F_4 + F_2 \wedge H_3 = 0,
\end{equation}
where $\tilde F_4, F_2, H_3$ are gauge invariant field strengths of rank $4,2,3$
respectively. 

Such relations complicate our usual understanding of charge in a gauge theory.
On the one hand, the fields $F_2$ and $H_3$ are invariant under
the gauge transformations naively associated with $\tilde F_4$ so that one would
not consider them to carry charge.  On the other, 
these fields are clearly sources of $\tilde F_4.$  Thus, one may ask what is the proper
definition of charge in a theory with Chern-Simons terms.  This question is central
to the issue raised by Bachas, Douglas, and Schweigert \cite{BDS} and continued
by several authors \cite{Taylor,JP,Mor} as to in just what 
sense D0-brane charge should be quantized.  

The approach adopted here is not to argue for a particular notion of charge, but instead
to discuss the fact that there are at least three natural notions of charge in a theory
with Chern-Simons terms or a modified Bianchi identity.
A closely related discussion in which multiple notions of charge were
of use can be found in \cite{IT}.  Which notion of charge
is most useful depends on the goal that one has in mind.  One of the main purposes
of this work is to provide a language for the proper discussion of these ideas.
The notions of charge discussed below are referred to as `brane source charge,'
`Maxwell charge,' and `Page charge.'  

Brane source charge is a notion of charge
most directly associated with external objects coupled to the theory.
As implied by the name, this charge is localized.  That is to say that it is
not carried by the gauge fields but is instead
associated directly with external sources (or topological non-trivialities
of the spacetime manifold) which take the shape of various branes.  This
charge is gauge invariant, but not conserved.  However, the non-conservation
rules take a precise form which 
can be directly related to the Hanany-Witten effect \cite{HW}.  The
relationship is a generalization of the argument for the case of
D0/D8-branes presented in \cite{PS,BGL}. 
In general, brane source charge is not quantized.
It is in fact this charge that was
directly computed by Bachas, Douglas, and Schweigert \cite{BDS}
and found not to be quantized in a particularly interesting example.
This is also the notion of charge used to identify the branes in
the supergravity solution of \cite{GM}.

Another notion, `Maxwell
charge,' is conserved and gauge invariant but not localized.   Instead, it
is carried by the gauge fields themselves and so is diffused throughout 
a classical solution.  As a result, the Dirac quantization argument
does not require its integral over an arbitrary volume to be quantized and, 
in general, it will be quantized only when integrated to infinity with
appropriate fall-off conditions on the fields.  It is this charge
that was recently discussed by Taylor \cite{Taylor}.

The third type of charge is ``Page charge.''  Here we follow
tradition (e.g., \cite{Stelle}) by naming this charge after the author of
the paper in which
it first appeared \cite{Page}.  This charge is again localized and not
carried by the gauge fields.  
It is also conserved and under appropriate conditions
it is invariant under small gauge transformations.  However, it does transform under
large gauge transformations.  By looking at how Chern-Simons terms
and modified Bianchi identities originate
in Kaluza-Klein reduction, one can argue that the Page charge is
quantized.  The Page charge quantization conditions were matched
with the Dirac quantization conditions of the higher dimensional theory
in \cite{BLPS}.
From the perspective of the theory on
the D2-brane, this is the charge that was conjectured to be quantized in
\cite{BDS} and, although it was not discussed in these terms there, it also
matches the notion of charge discussed by Alekseev, Mironov, and Morozov
in \cite{Mor}.

These types of charge are not
new, as they have all appeared in the literature.  However, as
is clear from the recent discussion of D0-brane charge in \cite{BDS,Taylor,Mor}, 
a coherent discussion of these charges will prove useful and a proper language
for discussing these charges is needed.  

We discuss in turn the brane source, Maxwell, and
Page notions of charge in sections II-IV.  Due to limitations of space, we discuss
the details only in the particularly illustrative case of D4-brane charge in type IIA
supergravity.
In each case, we make a number of observations about that particular notion of charge 
and the relation to D0-brane charge in the setting of Bachas, Douglas, and Schweigert.
A few closing comments are contained in section V.  

\section{Brane Source Charge and Brane-ending effects}
\label{bsSec}

Let us recall that that type IIA supergravity contains a three-form Ramond-Ramond gauge
field $A_3$ for which D4-branes carry magnetic charge.  One class of gauge transformations
act on this field as
$A_3 \rightarrow A_3 + d \Lambda_2$ for an arbitrary smooth 
two-form $\Lambda_2$.
Throughout this work, we find it convenient
to indicate the rank of each form with a subscript.
An unusual property of this field, however, is that it also transforms under
the gauge transformations normally associated with the Ramond-Ramond potential $A_1$:
\begin{equation}
(A_1,A_3) \rightarrow (A_1 + d \Lambda_0, A_3 - B_2 \wedge d \Lambda_0),
\end{equation}
where $B_2$ is the Neveu-Schwarz two-form (i.e., the Kalb-Ramond field).
This means that the field strength $F_4 =
dA_3$ is not gauge invariant, but instead transforms as $F_4 \rightarrow F_4 - H_3 \wedge
d \Lambda_0.$  Here, $H_3 = dB_2$ is the gauge invariant Neveu-Schwarz field strength.
As a result, it is convenient to introduce the gauge invariant `improved field strength'
$\tilde F_4 = dA_3
 - A_1 \wedge H_3$ and to write the Bianchi identity in the form
of equation (\ref{MBid}). Such a relation is known as a modified Bianchi identity.
Similar equations appear involving the dual field $*\tilde F_4$ (associated with D2-brane
charge) in the equations of motion due to Chern-Simons terms of the form $A_i \wedge F_j
\wedge F_k$ for various $i,j,k$ in the type IIA action.  One can often exchange
a modified Bianchi identity for a Chern-Simons term by performing an electromagnetic
duality transformation.  Due to their similar forms, our discussion in all cases below
applies equally well to the effects of modified Bianchi identities and those of Chern-Simons
terms.

We wish to discuss the various notions of charge in terms of a language of currents
associated with external sources.  This language, however, is sufficiently general
so as to be useful for what one
might call `solitonic charge' associated with topological
nontrivialities (such as black holes, 
any singularities that one might deem to allow, and so on).  Suppose for example
that we are given a spacetime containing a wormhole that is threaded by some electric
flux.  Then we may choose to consider a related spacetime
in which
the neck of the wormhole has been rounded off by hand.  The new spacetime will of course
not satisfy the supergravity equations of motion in the region that has been modified.
We can describe this departure from pure supergravity by saying that some external source
is present in this region.  Using such a language will allow us to suppose
that we work on the manifold $R^n$ and that the spacetime is smooth.

We begin with what, from the standpoint of the modified Bianchi identity, is perhaps the most
natural parameterization of this external source.  We simply define the nonvanishing
of the modified Bianchi identity to be the dual $*j_{D4}^{bs}$ of some current, which will
in some way be associated with D4-branes.  Thus, we have
\begin{equation}
\label{bs}
d \tilde F_4 + F_2 \wedge H_3 = *j_{D4}^{bs}.
\end{equation}
We repeat that this is nothing other than a definition of $*j_{D4}^{bs}$, now providing
a parameterization of the external sources.  In general, we would write each modified
Bianchi identity and equation of motion for the gauge fields 
as a polynomial in the gauge invariant improved
field strengths, their hodge duals, and exterior derivatives of these and then let
the right hand side be some $*j$.  Each such current will be associated with some brane, either
a D-brane, NS5-brane, or a fundamental string.  Similar sources for the metric are
associated with energy and momentum, while sources for the dilaton are associated
with NS instantons and NS7-branes.

Let us make a few simple observations about the current defined in (\ref{bs}).
Examining the left-hand side, we see that our current is gauge invariant.  It is also
`localized' in the sense that it vanishes wherever the spacetime is described by
pure supergravity.  In this sense, it is naturally associated with {\it external} brane
sources that are coupled to supergravity.  For this reason, we refer to this notion
of charge as `brane source charge.'  

We note that this notion of charge coincides with many familiar conventions.
For example, suppose that we rewrite type IIA supergravity in terms of the magnetic field
strength $A_5$ dual to $A_3.$  Then the modified Bianchi identity for $A_3$
becomes an equation of motion for $A_5$.  In this case, the brane source current is
just what results from additional terms of the form $-\int A_5 \wedge *j_{D4}^{bs}$ that
one would add to the action to represent external sources.
A similar discussion for the case of D0-brane charge on a D2-brane coupled to supergravity
shows that since brane source charge arises from varying the brane action
with respect to the gauge field, it is this 
notion of charge which raised the puzzle in
\cite{BDS}, as they found this charge not to be quantized.

In fact, supergravity considerations also lead one to expect this charge not
to be quantized.  This follows from the fact that it is not conserved, and that its
non-conservation takes a special form.  Let us simply take the exterior derivative
of (\ref{bs}), allowing also sources $*j_{D6}^{bs} = dF_2$ and $*j_{NS5}^{bs} = dH_3$ for
the other relevant gauge fields.  We find:
\begin{equation}
d*j^{bs}_{D4} = F_2 \wedge * j_{NS5}^{bs} + *j_{D6}^{bs} \wedge H_3,
\end{equation}
so that both NS5-branes and D6-branes can be sources of our charge in the proper backgrounds.
What is particularly interesting about this result is that, due to the ranks of the forms
involved, it has components in which all indices take spatial values.  This means
that such components have no time derivatives and instead constitute a {\it constraint}, telling
us how D4-brane charge must change in spatial directions.  In particular, 
integrating this result over some six-dimensional volume $V_6$ tells us that the net number of
D4-branes (as counted by brane-source charge) ending inside $V_6$ is controlled by
the fluxes of gauge fields captured by NS5-branes and D6-branes inside $V_6$:
\begin{equation}
\int_{V_6} *j_{D4} = \int_{V_6 \cap NS5} F_2 + \int_{V_6 \cap D6} H_3.
\end{equation}
Note that the intersection of $V_6$ with the worldvolume of an NS5-brane is generically
of dimension 2, and that the intersection with the worldvolume of a D6-brane is generically
of dimension 3.  The normalization is such that if a single NS5-brane captures all of the
$F_2$ flux emerging from a D6-brane, then this constraint states that exactly one
D4-brane worth of charge will begin (or end, depending on the sign) on the NS5-brane.
This constraint tells us that D4-brane source charge must be created continuously
over the world volume of NS5- and D6-branes.  Since constraints are typically not significantly
modified by quantization, it would be quite surprising if such a charge were quantized.
This point was also
made in \cite{BDS} working from the perspective of the worldvolume theory
on a brane.

Such constraints 
connect Chern-Simons terms and modified Bianchi identities with the same
types of branes ending on branes as in Townsend's `Brane Surgery' argument \cite{surgery}.
These arguments are not equivalent, however, as \cite{surgery} considers that case where
brane source charge (say, for a D4-brane)
is not created or destroyed, but instead flows away through the worldvolume of the other 
(D5- or D6-) brane.

Finally, we note (see also \cite{GM}) that such constraints provide
yet another 
derivation of the Hanany-Witten effect \cite{HW}.  The argument
is a generalization
of the argument of that of \cite{PS,BGL} for the D0/D8 case.
Suppose that an NS5-brane
lies on one side of a D6-brane in such a way that there is no D4-brane charge in the
vicinity.  Typically, the constraints can still be satisfied if other NS5- and D6-branes
are nearby.  When the NS5-brane is moved past the D6-brane, the flux captured by
each of these branes changes by one unit.  The NS5-brane must then be a source of one
D4-brane, while the D6-brane must be a sink.  If the branes are moved quickly, 
causality considerations show
that we must now have a D4-brane stretching from the NS5-brane to
the D6-brane.  Whether or not one wishes to use brane source charge to count
`real D4-branes,' one finds that some sort of brane must be stretched between the
NS5- and D6-branes.
A corresponding argument from the perspective of the worldvolume theory
was presented in \cite{DFK,K} 
but it is nice to arrive at this result via such a short argument
in supergravity.  Other complimentary derivations of this effect can be found in
\cite{Lif,dA,HoWu,OSZ,NOYY,Yosh}.  Some of these derivations use
an `anomaly inflow' argument, and we refer the reader to \cite{IT}
to connect such a perspective
directly with the present discussion,  closing the circle
of ideas.

\section{Maxwell Charge and Asymptotic Conditions}
\label{MaxSec}

Our next notion of charge follows from the idea that any source of the gauge
field should be considered to constitute a charge.  Consider again the relation
$d\tilde F_4 + F_2 \wedge H_3 = 0$ which holds in the absence of external sources.
Clearly, $F_2 \wedge H_3$ is a source for the field strength $\tilde F_4$, so that
we might count it as carrying charge.  To this end, let us define the Maxwell charge
current to be the exterior derivative of the gauge invariant field strength:
\begin{equation}
\label{Max}
d\tilde F_4 = *j_{D4}^{Maxwell}.
\end{equation}
Such a relation describes the familiar currents of Yang-Mills theories, in which the
gauge fields also carry charge.  A similar
idea allowing gravitational fields to contribute to energy and momentum
is captured by the ADM mass for gravity.  A study of \cite{Taylor} shows that 
this is in fact the notion of charge
used by Taylor in that reference.

This current has many useful properties.  It is manifestly gauge invariant
and conserved.  However, it is not localized, as it is carried by the bulk fields.
This means that the conservation law for Maxwell charge is somewhat less useful
that one might hope.  Consider for a moment integrating $\tilde F_4$ over some surface
$\partial V$ to obtain the total charge associated with some region $V$.  The charge
measured in this way is unchanged when we deform the surface $\partial V$ so long
as this surface does not pass through any charge.  Since Maxwell charge is carried by
the bulk fields, such charge-preserving deformations may not exist at all.

This of course is the case in Yang-Mills theory or gravity, where one solves the problem
by using Gauss' law for surfaces at infinity where the bulk charge density vanishes
under appropriate fall-off conditions.  This works well for charges carried by pointlike
objects, but is somewhat less satisfactory for the present case in which the sources
are branes.  The point is that one might like the charge measured to remain unchanged
when the Gauss' law surface is deformed in space as well as when translated in time.
A charge associated with $p$-branes is measured by a Gauss' law surface of co-dimension
$p+2$, so that interesting deformations of the Gauss' law surface in space are indeed
possible for $p > 0$.

Consider in particular the D4-brane case.  Note that the Maxwell and brane source
currents are related by $*j_{D4}^{Maxwell} = *j_{D4}^{bs} - F_2 \wedge H_3.$
Suppose that we have some region $V$ with $\partial V = S_1 - S_2$.  Then
$\int_{S_1} \tilde F_4 = \int_{S_2} \tilde F_4$ if and only if $\int_V *j_{D4}^{Maxwell} =0.$
In a region of infinity in which the supergravity equations of motion hold (and thus there
are no external sources), we have $\int_V *j_{D4}^{Maxwell} =  - \int_V  F_2 \wedge H_3.$
Note that this will not in general vanish (so long as $V$ spans a finite fraction
of infinity) as $\int F_2$ measures the D6-brane charge
while $\int H_3$ measures the NS5-brane charge.   The asymptotically
flat version of \cite{GM} or, analogously \cite{CGS} for D3-branes and test
D5-branes, are examples in which this
can be seen.  Note that one does not need the complete supergravity
solution to obtain this result.

Thus, even at infinity the Maxwell charge is not localized.
In fact, unless $F_2$ and
$H_3$ flux is confined,
the Maxwell D4
charge in a region $V$ must change continuously with $V$ even at infinity.
This means that Maxwell charge associated with generic surfaces at infinity cannot
be quantized.  Note, however, that in the case of D0-brane charge studied in
\cite{Taylor} there is a unique sphere at infinity at which Gauss' law can be applied
and the issue does not arise.

\section{Page Charge and Kaluza-Klein reduction}

The final notion of charge that we will consider is one
first introduced by Page in \cite{Page}.  The idea is first
to write the modified Bianchi identity (or equation of motion with
a Chern-Simons term) as the exterior derivative
of some differential form, which in general will not be gauge invariant.
In the presence of an external source, it is this exterior derivative
that is identified with a current or charge.  Thus, for our
case of D4-branes we would write
\begin{equation}
d( \tilde F_4 + A_1 \wedge H_3) = *j_{D4}^{Page}.
\end{equation}
There is some ambiguity in this process as the second term
could also have been taken to be of the form $F_2 \wedge B_2.$
This ambiguity will be discussed further below.

We see immediately that the Page current is conserved and
localized, in the sense that it vanishes when the pure supergravity
equations of motion hold.  However, it is also clear that this
current is gauge dependent as it transforms nontrivially under
gauge transformations of $A_1.$  This problem is to some extent
alleviated by integrating the current over some five-volume $V_5$ to
form a charge:
\begin{equation}
Q_{D4,V}^{Page} = \int_{V_5} *j_{D4}^{Page} = \int_{\partial V} (
\tilde F_4 + A_1 \wedge H_3).
\end{equation}
If $A_1$ is a well-defined 1-form on $\partial V$ and $dH_3=0$ on $\partial V$, 
then an integration by parts shows that the Page charge is invariant
under small gauge transformations $A_1 \rightarrow A_1 + d \Lambda_0$.
However, in general
it will still transform under large gauge transformations.
The qualification that $A_1$ be a well-defined 1-form means
that there can be no `Dirac strings' of $A_1$ passing through
$\partial V$ in the chosen gauge.  A similar integration by
parts shows that, when $\partial V$ does not intersect any
NS5 or D6 branes or the associated Dirac strings, the same page
charge would be obtained from $\tilde F_4 + F_2 \wedge B_2.$

We note that the Page charge differs from the Maxwell charge
only by the boundary term discussed in the last section.  That is, 
we have $Q^{Page}_{D4,V_5} = Q^{Maxwell}_{D4,V_5} + \int_{\partial V}
A_1 \wedge H_3.$  A similar expression holds for D0-brane charge.
For the case studied by Taylor in \cite{Taylor}, the corresponding
boundary term was explicitly assumed to vanish when $\partial V$
was the sphere at infinity.  Thus, although
\cite{Taylor} began with the idea of Maxwell charge, in that
case a discussion in terms of Page charge would be equivalent.
Similarly, when one works out the D0-brane Page charge for the case of
\cite{BDS} one finds $*j^{Page}_{D0} = *j_{D0}^{bs} -
\int B \wedge *j_{D2}^{bs}.$  It was exactly a term of the form
$\int B \wedge *j_{D2}^{bs}$ that created the puzzle in \cite{BDS}, and
we see that it is explicitly cancelled in the Page charge.  Computing
the Page charge for other examples agrees with \cite{Mor}, although it
was discussed there in a somewhat different language.

We would now like to argue that it is the Page charge which
is naturally quantized.  The argument that we will give is essentially
contained in \cite{BLPS} and perhaps earlier works as well.  
However, let us first embark on a small tangent
which is in fact not a convincing argument for quantization.
We note that D2-branes couple electrically to $\tilde F_4$ and
that the D2-brane action contains a term $\int_{D2} A_3.$
In order for $e^{iS_{D2}}$ to be insensitive to Dirac strings,
$\int_\Sigma A_3$ should be quantized for any 3-surface $\Sigma$ wrapping
tightly around a Dirac string.  But $\int_{\partial_V} (\tilde F_4
+ A_1 \wedge H_3) = \int_{\partial_V} (dA_3) = \int_\Sigma A_3$
where $\Sigma$ wraps tightly around all Dirac strings of $A_3$
passing through $\partial V$.  Thus, requiring $e^{iS_{D2}}$
to be well-defined in the presence of Dirac strings would
force quantization of the Page charge.  We agree with \cite{BDS}, however, 
that this is not by itself a convincing argument for quantization
of Page charge as it assumes that the effective action of the D2-brane is
known a priori.  In fact, the Chern-Simons terms of such an effective
action are typically deduced from properties of the bulk fields.
Nevertheless, it is reassuring that Page charge quantization is consistent
with the usual D2-brane action.

Now, for a more convincing argument.  Recall that
many of the Chern-Simons terms and modified Bianchi identities
of type IIA supergravity arise from the
Kaluza-Klein reduction of 11-dimensional supergravity.
Of course, 11-dimensional supergravity has its own Chern-Simons
terms as required by supersymmetry.  Nevertheless, our
discussion of D4-brane charge would be the same if, instead
of type IIA supergravity, we considered the reduction to
ten dimensions of an 11-dimensional Einstein Maxwell theory
given by 
\begin{equation}
S_{11} = \int \sqrt{g} R + \frac{1}{2} F^M_4 \wedge * F^M_4,
\end{equation}
and in particular having no Chern-Simons term.
We have labelled the 4-form field strength of this pseudo M-theory
$F_4^M$ in order to distinguish it from the $F_4$ of the ten
dimensional theory.

In such a simple Einstein-Maxwell theory, charge quantization
is believed to be well understood with $\int_{\partial V} F^M_4$ and
$\int_{\partial V} *F^M_4$ being quantized.  In Kaluza-Klein
reduction along $x_{10}$, the relation between 10- and 11-dimensional
fields is just
\begin{equation}
F^M_4 = F_4 + H_3 \wedge dx_{10} = (\tilde F_4 + A_1 \wedge H_3) + H_3
\wedge dx_{10}.
\end{equation}
As a result, if $Q_{D4}^{Page}(S_4) = \int_{S_4} (\tilde F_4 +
A_1 \wedge H_3)$ is the Page charge associated with the surface
$S_4$, we see that this is identical to the M5-brane charge
$Q_{M5}(S_4,x_{10}=const)$ defined by integrating $F_4^M$
over the surface at constant $x_{10}$ that projects to $S_4$
in the ten-dimensional spacetime.  This observation was used in \cite{BLPS} to match the ten-
and eleven-dimensional Dirac quantization conditions.  Thus, it is the Page
charge that lifts to the familiar notion of charge in 11-dimensions.
Quantization of the usual charge in 11-dimensional
Einstein-Maxwell theory directly implies quantization of D4-brane
Page charge in ten-dimensions.  It is for this reason that we have
chosen to use D4-brane charge as our example system.  Quantization
of the Page charge for other branes then follows from
T-duality.  T-duality directly implies Page charge quantization
in systems with sufficient translational symmetry, and one
can use homotopy invariance of the Page charge to complete
the argument. Quantization of the Page charge in 2+1 dimensional theories
with $A \wedge F$ Chern-Simons terms was derived in \cite{HT}.

Note that under the
Kaluza-Klein assumption of translation invariance in $x_{10}$ the
precise value of $x_{10}$ is unimportant.  Furthermore, under
a change of gauge $A_1 \rightarrow A_1 + d \Lambda_0$ in the
10-dimensional spacetime, we have $x_{10} \rightarrow x_{10} - \Lambda.$
This means that a change of gauge in ten dimensions corresponds to
a change of {\it surface} in 11-dimensions.  This provides
a clear physical meaning to the change in the Page charge under
a large gauge transformations: in the 11-dimensional theory, 
we have replaced the M5-brane charge contained in one surface
with the M5-brane charge contained in a homotopically
inequivalent surface.

\section{Discussion}
\label{Disc}

We have seen that three notions of charge can be useful
in theories with Chern-Simons terms.  Brane source charge
is gauge invariant and localized, but not conserved
or quantized.  Its non-conservation, however, summarizes
consistency conditions that must be satisfied by external
sources coupled to the theory and leads directly to the Hanany-Witten
brane creation effect.  

In contrast, Maxwell charge is
carried by the bulk fields and so is not localized.  It is quite similar
to the ADM mass, energy, and momentum of gravitating systems, which
is in fact one of the reasons for its use in \cite{Taylor}.  
This charge is both gauge invariant and conserved.  However, 
in certain interesting cases involving $p$-branes with $p>0$,
the fall-off conditions at infinity
are too weak for this conservation law to be as useful as one
might like.  

Finally, while it transforms nontrivially under large gauge transformations, 
the Page charge is localized and conserved.  When the Chern-Simons
term or modified Bianchi identity arises from Kaluza-Klein
reduction, this charge is naturally associated with charge in the
higher dimensional theory.  As a result, it is this charge that
is naturally taken to be quantized.  The gauge dependence
of the Page charge is nothing other than the ambiguity associated
with choosing a surface in the higher dimensional theory that
projects onto the chosen surface in the lower dimensional 
spacetime.
Note that, due to its
relation to the higher dimensional fields, it is also the
Page charge which is naturally associated with supersymmetry.

It is interesting to consider Page charge in the context of branes
created in the Hanany-Witten effect.  In many cases involving
D0- and D8-branes, 
the created string clearly has a Page charge of zero as the associated
Gauss' law surface can be slipped over the end of the D0-brane and
contracted to a point. However, a non-zero Page charge can arise for higher
branes.  A number of examples are under investigation.

Such considerations apply
not only to supergravity, but also for example to the D2-brane
theory directly investigated by Bachas, Douglas, and Schweigert.
They argued that a certain charge $\int F$ should be quantized, 
where $F=dA$ is a gauge field on the D2-brane that is in
fact not gauge invariant.  One can check that this is also
a Page charge of the D2-brane theory.  Again, Kaluza-Klein
reduction provides a useful perspective.  If one investigates
the relation between the D2-brane theory and the theory of an
M2-brane, one finds that $\int F$ is exactly the canonical
momentum of the M2-brane in the compact $x_{10}$ direction, and so is
again naturally quantized.

\acknowledgments

The author would like to thank Andr\'es Gomberoff, Rajesh Gopakumar, 
Michael Gutperle, 
Marc Henneaux, Rob Myers,
Djordje Minic, Shiraz Minwalla, Michael
Spalinski, Andy Strominger, Paul Townsend, and Arkady Tseytlin
for useful discussions.  This work was supported in part by
NSF grant PHY97-22362 to Syracuse University, 
the Alfred P. Sloan foundation, and by funds from Syracuse 
University.  

%%%%%%%%%%%%%%%%%%%%%%%%%%%%%%%%%%%%%%%%%%%%%%%%%%%%%%%%%%%


\begin{thebibliography}{99}
%%%%%%%%%%%%%%%%%%%%%%%%%%%%%%%%%%%%%%%%%%%%%%%%%%%%%%%%%%%

\bibitem{BDS} C. Bachas, M. Douglas, and C. Schweigert, {``Flux
Stabilization of D-branes,''} hep-th/0003037.

\bibitem{Taylor} W. Taylor, {\it ``D2-branes in B fields''}, 
hep-th/0004141.

\bibitem{JP} J. Polchinski, as cited in \cite{Taylor}.

\bibitem{Mor} A. Alekseev, A. Mironov, and A. Morozov, {\it ``On
B-independence of RR charges''}, hep-th/0005244.

\bibitem{IT} J.M. Izquierdo and P.K. Townsend, {\it ``Axionic Defect
Anomalies and their Cancellation,''} Nucl. Phys. {\bf B414} (1994)
93-113, hep-th/9307050.

\bibitem{HW} A. Hanany and E. Witten, {\it ``Type IIB
Superstrings, BPS Monopoles, and Three-Dimensional
Gauge Dynamics,''} Nucl. Phys. {\bf B492} (1997) 152, hep-th/9611230.

\bibitem{PS} J. Polchinski and A. Strominger, {``New Vacuua for Type 
II String Theory,''} Phys. Rev. Lett., {\bf B388} (1996) 736-742,
hep-th/9510227.

\bibitem{BGL} O. Bergman, M. R. Gaberdiel, and G. Lifschytz, 
{\it ``Branes, Orientifolds, and the Creation of Elementary
Strings,''}, Nucl. Phys. {\bf B509} (1998) 194-215, hep-th/9705130.

\bibitem{GM} A. Gomberoff and D. Marolf, {\it ``Brane Transmutation in
Supergravity,''} JHEP {\bf 02} (2000) 021.

\bibitem{Stelle} K. S. Stelle, {\it ``BPS branes in supergravity,''}
Trieste 1997, High energy physics and Cosmology,
hep-th/9803116.

\bibitem{Page} D. N. Page, {\it ``Classical stability of round
and squashed seven-spheres in eleven-dimensional supergravity''}
Phys. Rev. {\bf D 28},
2976 (1983).

\bibitem{BLPS} M.S. Bremer, H. Lu, C.N. Pope, and K.S. Stelle, 
{\it Dirac Quantization Conditions and Kaluza-Klein Reduction}, 
Nucl. Phys. {\bf B529} (1998) 259-294.

\bibitem{surgery} P.K. Townsend, {\it ``Brane Surgery''},
Nucl.\ Phys.\ Proc.\ Suppl.\ {\bf 58}
(1997) 163-175, hep-th/9609217.

\bibitem{DFK} U. Danielsson, G. Ferretti, and I. R. Klebanov,
{``Creation of Fundamental
Strings by Crossing D-branes,''} Phys. Rev. Lett. {\bf 79} (1997)
1984-1987, hep-th/9705084.

\bibitem{K} I. R. Klebanov, {``D-branes and Creation of Strings,''}
Nucl. Phys. Proc. Suppl. {\bf 68} (1998) 140,
hep-th/9709160.


\bibitem{Lif} G. Lifschytz, {\it Comparing D-branes to Black Branes},
hep-th/9604156.

\bibitem{dA} S. P. de Alwis, {``A note on brane creation,''}
Phys. Lett. {\bf B388} (1996) 720,
hep-th/9706142.

\bibitem{HoWu} P. Ho and Y. Wu, {\it Brane Creation in M(atrix)
Theory,''} Phys. Lett. {\bf B420} (1998) 43-50, hep-th/9708137.

\bibitem{OSZ} N. Ohta, T. Shimizu, and J-G Zhou, 
{\it ``Creation of Fundamental String in M(atrix) Theory,''}
Phys. Rev. {\bf D57} (1998) 2040-2044,
hep-th/9710218.

\bibitem{NOYY} T. Nakatsu, K. Ohta, T. Yokono, and Y.
Yoshia, {\it ``A proof of Brane Creation via M-theory,''}
Mod. Phys. Lett., {\bf A13} (1998) 293-302,
hep-th/9711117.

\bibitem{Yosh} Y. Yoshia, {\it ``Geometrical Analysis of Brane
Creation via $M$-theory,''} Prog. Theor. Phys., {\bf 99} (1998) 305-314,
hep-th/9711177.

\bibitem{CGS} C. G. Callan, A. Guijosa, and K. G. Savvidy, {``Baryons
and String Creation from the Fivebrane Woldvolume Action,''} Nucl. Phys.
{\bf B547} (1999) 127-142,
hep-th/9810092.

\bibitem{HT} M. Henneaux and C. Teitelboim, {\it ``Quantization
of Topological Mass in the Presence of a Magnetic Pole,}
Phys. Rev. Lett., {\bf 56} (1986) 689-692.

\end{thebibliography}
\end{document}